\documentclass[10pt]{article}
\usepackage{latexsym}
\usepackage{amsmath}
\usepackage{amsfonts}
\usepackage{graphicx}
\usepackage{url}
\usepackage{qic}

\def\calJ{{\cal J}}

\newcommand{\bea}{\begin{eqnarray} }

\newcommand{\eea}{\end{eqnarray} }

\newcommand{\ba}{\begin{array}}
\newcommand{\ea}{\end{array}}

\def\be{\begin{equation}}
\def\ee{\end{equation}}
\def\bea{\begin{eqnarray}}
\def\eea{\end{eqnarray}}
\newcommand{\ra}{\rangle}
\newcommand{\la}{\langle}
\newcommand{\ket}[1]{| #1\rangle}

\newtheorem{claim}{Claim}

\newtheorem{cor}[theorem]{Corollary}
\newtheorem{defn}[theorem]{Definition}
\newtheorem{thm}{Theorem}

\newcommand{\ignore}[1]{}

\newcommand{\trace}{\mathop{\mathrm{Tr}}\nolimits}
\newcommand{\RR}{\mathbb{R}}

\begin{document}

\title{Classical approximation schemes for the ground-state energy of quantum and classical Ising spin Hamiltonians on
planar graphs}

\author{Nikhil Bansal \thanks{IBM Watson Research Center, Yorktown Heights, NY, USA 10598.
\texttt{nikhil@us.ibm.com}} \and Sergey Bravyi \thanks{IBM Watson
Research Center, Yorktown Heights, NY, USA 10598.
\texttt{sbravyi@us.ibm.com}} \and Barbara M. Terhal \thanks{IBM
Watson Research Center, Yorktown Heights, NY, USA 10598.
\texttt{bterhal@gmail.com}}}

\maketitle

\begin{abstract}
We describe an efficient approximation algorithm for evaluating the
ground-state energy of the classical Ising Hamiltonian with linear
terms on an arbitrary planar graph. The running time of the
algorithm grows linearly with the number of spins and exponentially
with $1/\epsilon$, where $\epsilon$ is the worst-case relative
error. This result contrasts the well known fact that exact
computation of the ground-state energy for the two-dimensional Ising
spin glass model is NP-hard. We also present a classical
approximation algorithm for the Local Hamiltonian Problem or Quantum
Ising Spin Glass problem on a planar graph {\em with bounded degree}
which is known to be a QMA-complete problem. Using a different
technique we find a classical approximation algorithm for the
quantum Ising spin glass problem on the simplest planar graph with
unbounded degree, the star graph.
\end{abstract}

\section{Introduction}

Ising spin glasses model physical spin systems with random,
competing interactions due to disorder in the system \cite{book:FH}.
In order to make meaningful predictions about such systems one can
consider statistical ensembles that represent different realizations
of the couplings. 
For a particular realization of the couplings
%
one is generally interested in finding algorithms to determine
properties such as the spectrum, partition function, the
ground-state or the ground-state energy. Algorithms with a running
time that is a polynomial in the problem size are called {\em
efficient}, in contrast with inefficient procedures that take
super-polynomial or exponential running times. Connections between
disordered systems in statistical physics and questions of
computational complexity have been extensively explored, see e.g.
\cite{FA:statNP} and \cite{KS:statcc2}.

In this paper we find a new application of computational complexity
tools to spin glass problems, namely a rigorous approximation
algorithm to determine the ground-state energy of a classical or
quantum Ising spin glass on a planar graph. It has been shown that
to determine the ground-state energy of a classical Ising spin glass
on a 2D lattice with linear terms
 {\it exactly}
is computationally hard, or NP-complete \cite{barahona:np}. Terhal
and Oliveira showed that determining the smallest eigenvalue of a
quantum Ising spin glass of $n$ qubits on a planar graph with
$1/{\rm poly}(n)$ accuracy is QMA-complete \cite{OT:qma}. Aharonov
{\em et al.} \cite{AGIK:1d} showed that even determining the
smallest eigenvalue for {\em qudits} on a line
is QMA-complete.
Therefore the running time of any algorithm for computing the ground
state energy exactly must scale super-polynomially with the number
of spins $n$ (under the assumption P$\ne $NP). Thus it is natural to
look for approximation algorithms that solve the problem in
polynomial time at the cost of providing a slightly non-optimal
solution.

Let us define the (quantum) Ising spin glass problem precisely. Let
$G=(V,E)$ denote any graph with $n$ vertices and let $u,v \in V$ be
vertices of the graph. One can associate a $2$-local Ising spin
glass Hamiltonian $H$ with interaction graph $G=(V,E)$, \be
H=\sum_{(u,v)\in E} Q_{u,v} + \sum_{u\in V} L_u=Q+L.
\label{eq:defqH}\ee Here $Q_{u,v}$ is quadratic in Pauli operators
and $L_u$ is linear. We assume that $||Q_{u,v}||,||L_u|| \leq {\rm
poly}(n)$. It is also assumed that ${\rm Tr}\, H=0$. We obtain the
{\em classical} Ising spin glass, the function $H(S)$, by setting
$Q_{u,v}=c_{uv} S_u S_v$ and $L_u=d_u S_u$ for classical spins
$S_u=\{-1,1\}$, i.e. \be H(S)=\sum_{u,v} c_{uv}S_u S_v+\sum_u d_u
S_u \label{eq:defcH} \ee In the classical case $c_{uv}$ and $d_u$
can be given as some $m$-bit numbers. Bieche et al. \cite{bmru80}
has shown that for planar graphs $G$ the problem of determining the
minimum value of $H(S)$ (the ground-state energy) and the associated
assignment $S$ can be solved efficiently if there are no linear
terms (i.e. all $d_u=0$).

%

Our approximation algorithm for the classical and quantum Ising spin
glass is relevant in light of the recent research on quantum
adiabatic approaches for finding the ground-state of a classical or
quantum Ising spin glass and solving other NP-complete problems. The
paradigm of adiabatic quantum computation
 (AQC) was first introduced in
\cite{FGGS:adia}. An adiabatic computation proceeds by slowly
varying the system's Hamiltonian starting from some simple
Hamiltonian $H_{0}$ at the time $t=0$ and arriving to, for example,
a classical Ising spin glass Hamiltonian $H$ at $t=T$ (the final
Hamiltonian can also capture other NP-hard problems). The adiabatic
theorem, see e.g. \cite{AR:adia}, guarantees that if one starts from
the ground state of $H_{0}$ and the running time $T$ is large
compared to the inverse spectral gap at all times then the final
state is close the ground-state of $H$.

If we assume the validity of the conjecture that a quantum computer
can not solve NP-complete nor QMA-complete problems (see for
classical spin glasses the analysis
in~\cite{DMV:adia,FGGN:adiafail})
 it follows that
AQC is a means to obtain an {\em approximation} to the ground-state
and the ground-state energy. The quality of this approximation and
its dependence on the hardness of the problem and the adiabatic path
are at present not well understood. Some experiments on a physical
realization of the classical Ising model in a transverse field show
that quantum annealing can lead to faster equilibration of the
system \cite{brooke+}. Also, it was recently shown that a quantum
algorithm can provide a square-root speed-up over a classical
simulated annealing algorithm \cite{somma+}. These results do not
show that the Ising spin glass problem becomes easy on a quantum
computer, but point to possible advantages of using a quantum computer to obtain
efficient approximation algorithms. Recently the company D-wave has claimed to have implemented
the Ising spin glass
Hamiltonian
 (with additional edges on the diagonals of the lattice)
and an adiabatic evolution for $16$ qubits on a $4 \times 4$ square
lattice, see \cite{dwave:announce}. The hope of this endeavor is
that such system outperforms classical computers in (approximately)
solving optimization problems.

Given these claims and results about quantum speed-ups it is clearly
interesting to consider how well an approximation to the
ground-state energy can be obtained by purely classical algorithmic
means.

The area of approximation algorithms is an active area of research
in computer science, see e.g. \cite{book:vazirani}.
Such algorithms are often of practical importance for generic hard
problems for which we are willing to trade off the relative quality
of the approximation versus the running time of the algorithm. Three
main types of approximations to optimization problems can be
distinguished. A problem is said to have a polynomial time
approximation scheme (PTAS) if given any $\epsilon >0$, there is an
algorithm $A_\epsilon$ which for any instance $I$ produces a
solution within
$(1\pm \epsilon)$ times the optimal solution. In addition
$A_\epsilon$ has a running time which is a polynomial in the input
size of $I$. Observe that the running time of a PTAS is only
required to be polynomial in input size, and it can have an
arbitrary dependence on $\epsilon$ (for example $n^{1/\epsilon}$ or
$2^{1/\epsilon}$ poly$(n)$ are valid running times for a PTAS). A
stronger notion is that of a fully polynomial time approximation
scheme (FPTAS), where the running time of the approximation scheme
is required to be polynomial both in the size of the input and in
$(1/\epsilon)$. For the classical Ising spin glass problem an
instance is a particular graph and set of values of the weights
$c_{uv}$ and $d_u$. The optimal solution is the minimum value of the
energy $H(S)$.

For the Ising spin glass problem on planar graphs,
we can exclude the possibility of a FPTAS (assuming ${\rm P} \neq
{\rm NP}$). This is due to the fact that the problem is NP-hard even
when $c_{uv},d_u$ are restricted to be either $-1,0$ or $+1$
\cite{barahona:np}. Let us assume that we have a FPTAS and set
$\epsilon=\delta/{\rm poly}(n)$ for some constant $\delta$. This
gives a polynomial-time algorithm to approximate $\min_S H(S)$ with
an error which is at most $\delta \min_S H(S)/{\rm poly}(n)$. This
is sufficient accuracy to solve the NP-complete problem {\em
exactly} since $H(S)$ is at most ${\rm poly}(n)$ and the difference
between the minimum of $H(S)$ and the value right above it (i.e. the
energy gap) is at least $1$, that is, independent of $n$.

A class of classical optimization problems for which the objective
function can be efficiently approximated with a relative error
$\epsilon$ for some {\it fixed } $\epsilon$ is called APX. It is
known that some problems in APX do not have a PTAS (under the
assumption P$\ne$NP), for example, Minimum Vertex Cover and Maximum
Cut problems, see~\cite{AK:apx-hard}.
 In other words, for such problems no polynomial-time algorithm can make the relative error smaller
 than some constant threshold value $\epsilon_0$.
 Problems having this property are called APX-hard.
One can use the relation between the Ising spin glass problem and the
Maximum Cut problem, see~\cite{JS:ising}, to show that the former
is APX-hard if defined on arbitrary graphs. This is the reason why the present paper focuses only on planar graphs.

Note that the existence of a PTAS for a Hamiltonian $H$ does not
imply the existence of a PTAS for the trivially related problem of
finding the ground-state energy of
$H+a I$, where $a$ is a constant; this is because the PTAS produces
a solution with small {\em relative} error. Our PTAS for the quantum
and classical Ising spin glass problem can be translated to an
approximation algorithm with an {\it absolute} error at most
$\epsilon W$, where $W=\sum_{(u,v)} ||Q_{uv}||$ (see e.g.
Eq.~(\ref{error}) for the classical error analysis). Clearly, while
comparing the quality of an approximation obtained using AQC and the
classical PTAS the relevant figure of merit must be an absolute
error, because the Hamiltonians $H$ and $H+aI$ are physically
equivalent.
Note that when we state our running times for the approximation algorithms, we state the
{\em worst-case} running time depending on some guaranteed error-bound. In practice, running times
may be much faster if heuristic methods are used within the approximation algorithm.
Such heuristic methods cannot guarantee an error-bound, but may work well
for average-case or `real-life' instances.

We will first consider the classical Ising spin glass on a graph $G$
which is a two-dimensional lattice, and give a PTAS for this case.
It has a running time $T=O(n4^{\frac1\epsilon})$, see Section
\ref{sec:2d}.
The intuitive idea behind this construction is
simple. Assume, for simplicity, that all couplings between spins
have similar
 strength. Then one can subdivide a 2D lattice into blocks of size
 $L \times L$ by omitting the edges connecting these subblocks. The
 total contribution of these omitted boundary edges scales as
 $4L \times \frac{n}{L^2}=O(n/L)$ and hence for large, but constant, $L=1/\epsilon$ the
 error that one makes by omitting these edges is bounded by at most $O(\epsilon
 n)$. This proves that there exists an approximation algorithm with absolute error.
However one can show that the ground-state energy scales with $n$
(see e.g. the rigorous Theorem \ref{th:main}) and thus the error
will be proportional to the ground-state energy which is exactly
what is desired for the PTAS.

 In our formulation of the problem, not all edges on the 2D lattice have similar strength,
 hence somewhat more elaborate arguments are needed to show the existence of
 a PTAS.

For general planar graph this situation is more involved. Vertices
in the graph can have arbitrary high degree and it is not clear how
to divide up the graph into sub-blocks with small boundaries.
Let $W=\sum_{(u,v)} ||Q_{u,v}||$ for the quantum Isin spin glass and
$W=\sum_{(u,v)} |c_{uv}|$ for the classical Isin spin glass, see the
definitions of the Hamiltonians in Eqs.
(\ref{eq:defqH},\ref{eq:defcH}). In the `classical' Theorem
\ref{th:main} and its quantum counterpart, Theorem \ref{thm:extq}, we
will show that the ground-state energy of an Ising spin glass on a
planar graph is less than $-c W$ for some constant $c$. This
rigorously expresses the intuitive physical notion that the
ground-state energy is extensive in the system size $n$. The idea of
the PTAS is then as follows. We take out a subset of edges in the
planar graph for which (i) one can show that they contribute at most
$\epsilon W$ to the Hamiltonian and (ii) by taking out these edges
one ends up with a set of simpler disconnected graphs on which one
can solve the problem efficiently (in ${\rm
poly}(n)2^{O(1/\epsilon)}$ time). Let $\tilde{H}$ denote the
(classical or quantum) Hamiltonian for the problem where we have taken
out these edges. Since we only take out edges (and no vertices), a
state with minimum energy for $\tilde{H}$ is also a state for $H$.
Let $\lambda(H)$ denote the ground-state energy of $H$. By Weyl's
inequality $|\lambda(H)-\lambda(\tilde{H})| \leq ||H-\tilde{H}||
\leq \epsilon W$. Using Theorems \ref{th:main} and \ref{thm:extq} we
can relate $W$ back to the lowest eigenvalue of $H$ and hence show
that the PTAS outputs a (quantum) state which has energy at most
$O(\epsilon \lambda(H))$ higher than the true ground-state energy.

How do we take out edges from the original planar graph? In the
classical case, see Section \ref{sec:approx}, we take out edges
related to a so-called outerplanar decomposition of a graph. In this
way we end up with disconnected graphs which have a constant {\em
tree-width}. It is known how to solve the classical spin problem on
graphs with bounded tree-width (using dynamic programming).

In the quantum case we cannot chose this procedure since the quantum
problem on a graph with bounded-tree width, or even on a tree, can
still be hard, see \cite{AGIK:1d}. This points to an interesting
difference between the quantum and the classical Isin spin glass.

If, in the quantum case, we restrict ourselves to graphs with
bounded-degree, we can apply a procedure that removes edges and
leaves a set of disconnected graphs each of which has {\em constant}
size (related to $\epsilon$), see Section \ref{sec:qapprox}.
Determining the ground-state energy of a Hamiltonian in a space of constant
dimension can be done classically. Note that this PTAS
outputs a classical description of a quantum state which has an
energy $O(\epsilon \lambda(H))$-close to the true ground-state
energy.

Our last result finds a PTAS for the quantum problem on a star
graph, see Section \ref{sec:qstar}. The complexity of the local
Hamiltonian problem on this graph is not known. For the star graph
it is not clear how to take out edges without introducing a
large error. Hence we will use a different technique which uses
symmetry and the rounding of interactions. In effect, we construct a
Hamiltonian $\tilde{H}$ for which $\lambda(\tilde{H})$ and its
ground-state can be determined efficiently {\em and}
$||H-\tilde{H}|| \leq \epsilon W$. The construction works as long as
all terms in $H$ have norms in a range $[a,1]$ for a constant $a$.
Such condition was not present in the other PTAS constructions.
Extensions of this technique may be important for addressing the
general quantum problem on planar graphs.

We note that our technique for the classical planar graph problem,
i.e. using an outerplanar decomposition of the graph, is fairly
standard for solving hard classical problems on planar graphs. In
fact, many problems admit a PTAS on planar graphs even though
approximating them on general graphs is known to be NP-hard, see
\cite{LT80,bak94,kha96}. While our techniques are similar to those
of \cite{bak94} and \cite{kha96} at a conceptual level, our results
do not follow directly from their work and require some new ideas.
The main difficulty is that the Hamiltonian involves both positive
and negative terms which can possibly cancel out.

\section{Classical Hamiltonians on planar graphs}
\label{sec:planar}
\subsection{The 2D Lattice Case}
\begin{figure}[htb]\centerline{\includegraphics[scale=.20]{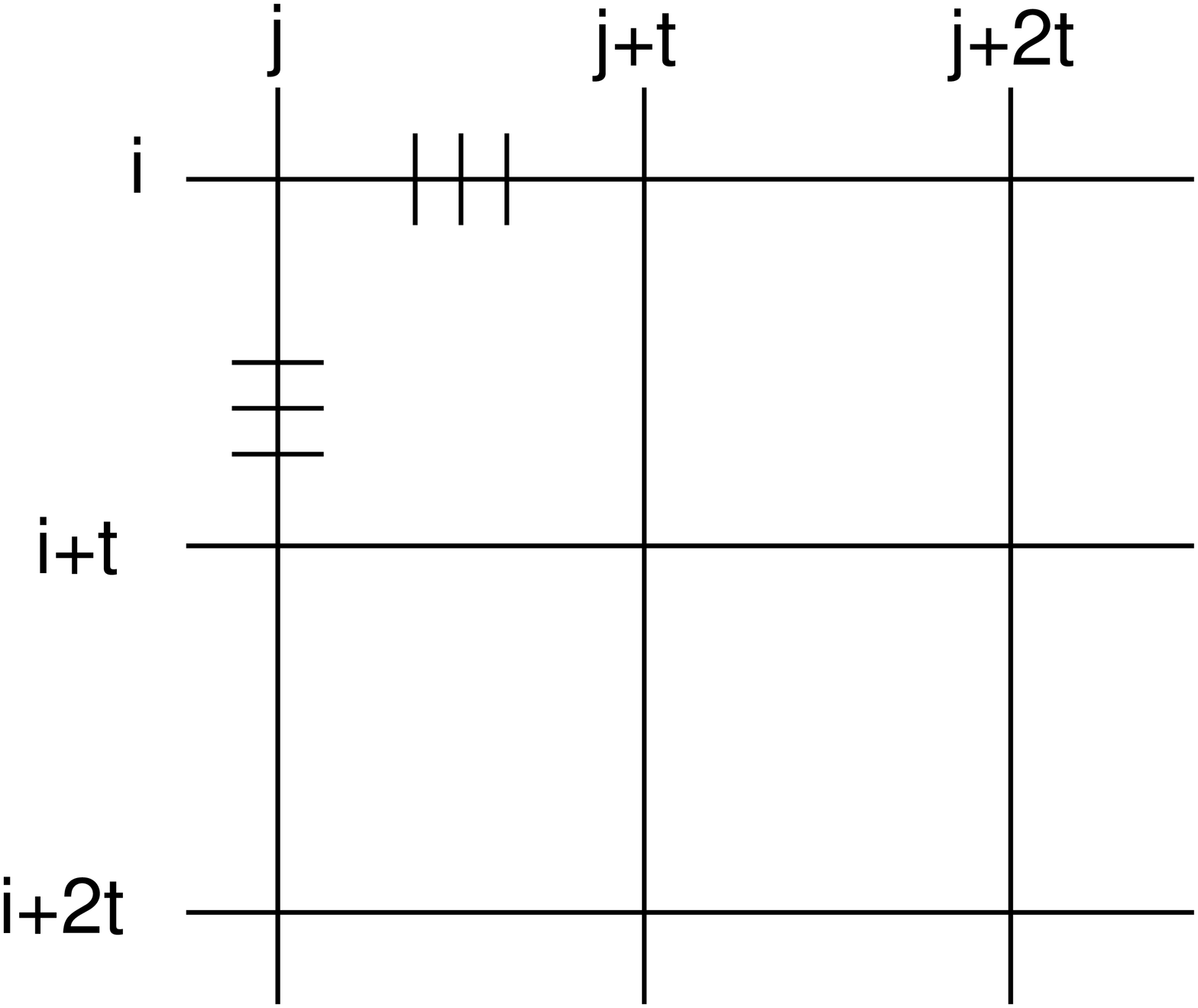}}
\fcaption{Sets of vertices on horizontal lines $X_i$ and sets of
vertices at vertical lines $Y_j$. Drawn are also some vertical edges
that are part of $H_i^x$ and horizontal edges that are part of
$H_j^y$.\label{fig:sets}}\end{figure}

Consider the classical Ising spin glass Hamiltonian Eq.~(\ref{eq:defcH})  defined on
a 2D square lattice of size $\sqrt{n}\times \sqrt{n}$.
\label{sec:2d} Let $\epsilon>0$ be a fixed small constant. Without
loss of generality let us assume that $t= 1/\epsilon$ is an integer.
For $i=0,\ldots,t-1$, let $X_i$ denote the set of vertices $u=(x,y)$
on the horizontal lines defined by \{$y \equiv  i$ (modulo $t$) \}.
Similarly, let $Y_i$ denote the set of vertices on the vertical
lines  defined by \{ $x \equiv i$ (modulo $t$) \}, see Figure
\ref{fig:sets}. We define the Hamiltonians $H_i^x(S)$ and $H_j^y(S)$
as follows. $H_i^x$ includes all terms $d_u S_u$, $u\in X_i$ and all
terms $c_{uv} \, S_u S_v$ such that $(u,v)$ is a vertical edge that
has exactly one end-point in $X_i$. Similarly $H_j^y$ includes all
terms $d_u S_u$ for $u \in Y_j$ and and all terms $c_{uv} \, S_u
S_v$ such that $(u,v)$ is a horizontal edge that has exactly one
end-point in $Y_j$. It is easy to see that
 $\sum_{i=0}^{t-1} H_i^x(S) +
H_i^y(S)=2H(S)$\footnote{We note that a similar decomposition could
be obtained for qubits on the square lattice with additional
diagonal interactions. In such case we would define four
Hamiltonians corresponding to lines of vertical, horizontal,
diagonal-$\backslash$, and diagonal-$/$ vertices with the edges
incident on these vertices.}. This implies that there exists an $i$
and $b=x$ or $y$ such that $H_i^b(S_{\rm opt}) \geq H(S_{\rm
opt})/t$ or $H(S_{\rm opt})-H_i^b(S_{\rm opt}) \leq (1-\epsilon)
H(S_{\rm opt})$,
where $S_{{\rm opt}}$ is a spin configuration with the minimum
energy.

For any $i$ and $b=x,y$ consider a Hamiltonian $H_{sub,i}^b(S) =H(S)-H_i^b(S)$.
Note that this Hamiltonian describes the Ising model defined on a set of $\epsilon\sqrt{n}$ disconnected strips
of size $\frac1\epsilon \times \sqrt{n}$ and $\epsilon \sqrt{n}$ disconnected  lines of size $\sqrt{n}$.
The latter corresponds to sets of edges that have both endpoints in $X_i$ or $Y_i$ and thus contains no linear terms.
It means that $H_{sub,i}^b(S)$ is invariant under flipping all the spins in any connected component
of $X_i$ (if $b=x$) or $Y_i$ (if $b=y$). Besides, as shown in the previous paragraph, there exists
a choice of $i$ and $b$ such that $H_{sub,i}^b(S_{opt})\le (1-\epsilon) H(S_{opt})$.

Let $S_{opt}'$ be  a spin configuration that achieves the minimum of $H_{sub,i}^b(S)$ for some
fixed $i$ and $b$.
Note that $S_{opt}'$ assigns values to all the vertices of the lattice. Using the symmetry of $H_{sub,i}^b(S)$
mentioned above, one can choose $S_{opt}'$ such that $H_i^b(S_{opt}')\le 0$.  Indeed, if $b=x$ then
$H_i^x(S_{opt}')$ changes a sign under flipping all the spins in $X_i$ while $H_{sub,i}^x(S)$ is invariant
under this flip (the same argument applies to $b=y$). Therefore,
for any $i$ and $b$ one has $H(S_{opt})\le H(S_{opt}')\le  H_{sub,i}^b(S_{opt}')$ and
for some $i$ and $b$ one has $H_{sub,i}^b(S_{opt}') \le H_{sub,i}^b(S_{opt})\le (1-\epsilon) H(S_{opt})$.
Thus the minimum energy of $H_{sub,i}^b(S)$ over all $i$, $b$, and $S$
approximates $H(S_{opt})$ within a factor $1-\epsilon$.

It follows that we can get a PTAS by finding the minimum of
$H_{sub,i}^b(S)$ over all choices of $i=0,\ldots,t-1$, $b=x,y$, and
all spin configurations $S$. The running time of this PTAS is
$T=\frac{2}{\epsilon} \cdot \left[
T_{strip}(\sqrt{n},\frac1\epsilon)+T_{strip}(\sqrt{n},1) \right]
\cdot O(\epsilon \sqrt{n})$, where $T_{strip}(r,s)$ is the running
time needed to find the optimal solution for the Ising spin glass
Hamiltonian on a $r\times s$ strip. One can easily show that

\begin{lemma}
There is a dynamic programming algorithm that computes the optimum
solution for a $r \times s$ lattice-blocks using space $O(2^s)$ and
time $O(r 4^s)$.
\end{lemma}

{\bf Proof}: Let $B_i$ denote the lattice $B$ restricted to row $i$
or less ($B_r = B$). Let $S$ be  $\{-1,+1\}$ vector of size $s$ and
for $i \in \{1,\ldots,r\}$, let $V(i,S)$ be the optimum value of the
Hamiltonian restricted to $B_i$ such that the variables on the
$i$-th row have assignment $S$. The dynamic program  computes and
stores $V(i,S)$ for all $i$ and $S$ starting sequentially from
$i=1$. This suffices as the optimum solution for $B$ is exactly
$\min_S V(r,S)$.

For $i=1$, the quantity $V(1,S)$ can be easily computed for each
$S$. Suppose  $V(i,S)$  has been computed and stored for all $S$.
For an assignment $S$ of row $i+1$ and an assignment $S'$ of row
$i$, let $Z(i+1,S,S')$ denote the contribution of all terms to the
Hamiltonian corresponding to vertices on row $i+1$, and edges in
$B_{i+1}$ with at least one end point in row $i+1$. Since the
assignment $S$ on row $i+1$ can only affect $Z$, it follows that \be
V(i+1,S) = \min_{S'} ( V(i,S') + Z(i+1,S,S')). \label{minf} \ee
Since $Z(i+1,S,S')$ can be computed in time $O(s)$ and $V(i,S')$ are
already stored, computing  $V(i+1,S)$ for all $S$ takes time $O(s2^s
\cdot 2^s)$. We can speed up this procedure somewhat to $O(4^s)$ by
considering the assignments $S'$ in  say the Gray code order (where
successive assignments differ in exactly 1 variable), and hence only
$O(1)$ work needs to be done per assignment $S'$.
\begin{flushright}
$\Box$
\end{flushright}

Thus using the dynamic program leads to a PTAS with an overall running time $T=O(n4^{\frac1{\epsilon}})$.

\subsection{Quantum Speed-Up?}
\label{sec:qspeed}

Let us consider how a quantum computer could improve these running
times. A possible application of quantum searching is inside the
dynamic programming. Since many classical approximation algorithms
rely on dynamic programming, this could be an important area of
applications for quantum searching. For each row of $s$ spins the dynamic program performs
$2^s$ minimizations and the minimum is over a function which takes
$2^s$ values. There exists a quantum algorithm for finding the
minimum of a function \cite{DH:minimum} which provides a square-root speed-up
over a brute-force classical minimization. This algorithm uses the
Grover search algorithm as a subroutine. However in the dynamic
program, part of the input to this function is {\em stored} which
implies that this is a problem of searching in a real database. In
the database setting one has to consider the additional
time/hardware overhead in accessing the spatially extended database.
Optical or classical wave implementations of this type of searching
have been considered, see \cite{KMSW:opt_grover}.
Whether this application of Grover's algorithm in dynamic programming is of genuine interest
will depend how its physical implementation competes in practice
with the capabilities of classical computers.

An alternative, less efficient, way to divide up the lattice is to
remove edges that connect subblocks of $L \times L$ spins. This
leads to a classical approximation algorithm with running time
$O(2^{O(1/\epsilon^2)})$. In this scenario, a direct quantum search
algorithm can help in finding the minimum energy of each block, but
note that this quadratic improvement would still lead to a running
time which scales as $2^{O(1/\epsilon^2)}$ which is {\em worse} than
the dynamic programming method.

\subsection{General Planar Graphs}
Let us now consider the general case of planar graphs. In the
lattice case we used the symmetry of the lattice to argue that there
exists a small subset of edges such that they have a small
contribution to the Hamiltonian and removing them decomposes the
lattice into small disjoint blocks. For general planar graphs we
cannot use this argument due to the lack of symmetry and hence we
argue indirectly.
We show that the magnitude of the optimal solution is at least a constant fraction
of the sum of the absolute values of quadratic terms corresponding to the edges. This allows us
to find a subset of edges with relatively small weight such that removing them decomposes
the graph into simpler disjoint graphs for which the problem can be solved directly.
Since the removed edges have small weight adding them back in does not increase the
Hamiltonian by too much, irrespective of
whether they are satisfied or not.

\subsection{Main theorem and Its Proof}

For a planar graph $G=(V,E)$ let $W = \sum_{(u,v) \in E} |c_{uv}|$.
The key to our PTAS is the following result that shows that the
value of the optimal assignment scales linearly with $W$:

\begin{thm}
\label{th:main} If $G$ is planar, then $H(S_{\rm opt}) \leq -W/3$.
\end{thm}

We begin with a simple property of the optimal solution (the one
that minimizes $H(S)$) that holds for an arbitrary graph.
Let us write the Hamiltonian as $H(S)=Q(S)+L(S)$, where $Q(.)$ is
the contribution of the quadratic (2-local) terms and $L(.)$ is the
contribution of the linear (1-local) terms.
\begin{claim}
\label{cl1} There exists an optimal solution $S_{opt}$ such that
$H(S_{\rm opt}) \leq 0$ and $L(S_{\rm opt}) \leq 0$.
 \end{claim}

{\bf Proof}: It is clear that $\sum_S H(S)=0$ (this is the traceless
condition). This implies that there must exist a spin configuration
with negative energy and thus $H(S_{\rm opt})$ is negative. The
second part can be argued by assuming the contrary ($L(S_{\rm opt})>
0$) and then noting that the solution with opposites signs $-S_{\rm
opt}$ is better than $S_{\rm opt}$ itself. $\Box$

Thus we note that it suffices to show that $\min_S Q(S) \leq -W/3$
since Claim \ref{cl1} shows that $\min_S H(S) \leq \min_S Q(S)$.
Hence we consider planar graphs with only quadratic terms in $H(S)$,
i.e. we assume that $d_u=0$ for all $u$. Recall that Bieche et al.
\cite{bmru80} has shown that this problem can be solved exactly in
polynomial time. Our proof of Theorem \ref{th:main} builds on the
ideas of Bieche and so we first describe these ideas. We begin with
some notation.

A graph is planar if it can be drawn in the plane such that no edges
cross. This drawing defines disjoint regions in the plane that are
called faces. A cycle in a graph is a collection of edges
$(u_1,u_2), (u_2,u_3),\ldots, (u_{\ell-1},u_\ell)$ where $u_\ell =
u_1$. Given two cycles $C_1$ and $C_2$, their sum $C_1 \oplus C_2$
is defined as the symmetric difference of $C_1$ and $C_2$. The faces
of planar graph form a cycle basis, that is, every cycle can be
expressed as a sum of faces. Given an assignment $S$, an edge
$(u,v)$ is called unsatisfied if $u$ and $v$ are not assigned
according the sign of $c_{uv}$ (i.e. if $c_{uv}  \geq 0$ but $S_uS_v
= 1$ or  if $c_{uv} <0$ but $S_uS_v = -1$). A face $F$ is called
frustrated if it contains an odd number of edges with positive
weight. A key observation is that for any assignment, a frustrated
face must always contain an odd number (hence at least one) of
unsatisfied edges. Conversely, if $J$ is a subset of edges such that
each frustrated (resp. non-frustrated) face contains exactly an odd
(resp. even) number of edges in $J$, then there is an assignment $S$
such that the unsatisfied edges are exactly those in $J$. Thus,
$Q(S) = -W + 2 \sum_{(u,v)\in J}| c_{uv}|$. Bieche et al.
\cite{bmru80} showed that for planar graphs finding such a set $J$
with minimum weight is equivalent to finding the minimum weight
$T$-join in the dual graph $G^*$ (these terms are defined below).
Let us remark that the total number of frustrated faces is even because
each edge $(u,v)$ with $c_{u,v}>0$ has exactly two adjacent faces.

For a planar graph $G$, its dual graph $G^*$ is defined as follows:
$G^*$ has a vertex for each face in $G$. Vertices $u$ and $v$ in
$G^*$ are connected by an edge if and only if the faces
corresponding to $u$ and $v$ in $G$ share a common edge. Given a
$G$, the dual $G^*$ is not necessarily unique (it depends on the
drawing of $G$), however $G^{**} = G$.
A subset of edges $E'$ is
called a cut-set if removing them disconnects the graph into two or
more components. For planar graphs, every cut-set in the dual graph
$G^*$ corresponds to a cycle in $G$.
We shall see below that sets of unsatisfied edges $\calJ$ in $G$ can be identified
with $T$-joins in the dual graph $G^*$, where $T$ is a set of frustrated faces.

Let $G=(V,E)$ be a graph with
edge weights, and let $T$ be a subset of vertices $T \subseteq V$
such that $|T|$ is even.  A {{\em $T$-join} is a collection of edges
$J$ such that each vertex in $T$ is adjacent to an odd number of
edges in $J$ and  each vertex in $V\setminus T$ is adjacent to an
even number of edges in $J$. The minimum weighted $T$-join problem
is to find a $T$-join with minimum weight, and this can be found in
polynomial time using matchings. As we mentioned above finding the
optimal assignment is equivalent to finding the minimum weight
$T$-join in  $G^*$ where $T$ to be the set of vertices corresponding
to the frustrated faces in $G$, and where an edge $e$ corresponding to
$(u,v)$ in $G$ has weight $w(e)=|c_{uv}|$.

We will use a polyhedral description of $T$-joins. For a subset of
edges $J$, let $v(J)$ denote the corresponding $|E|$-dimensional
incidence vector (with 1 in the $i$-th coordinate if edge $i$ lies
in $J$ and 0 otherwise). For a subset of vertices $X$, let
$\delta(X)$ denote the set of edges with one end point in $X$ and
other in $V\setminus X$. Given a graph $G$ and the set $T$, we say
that a subset of edges $J$ is an upper $T$-join if some subset $J'$
of $J$ is a $T$-join for $G$. Let $P$ be the convex hull of all
vectors $v(J)$ corresponding to the incidence vector of upper
$T$-joins. $P$ is called the up-polyhedra of $T$-joins. Edmonds and
Johnson \cite{EJ73} gave the following exact description of $P$ (see
the book by Schrijver \cite{sch03}, Chapter 29, par. 1-6 for further
details).
\begin{eqnarray}
\sum_{e \in \delta(W)} x(e) \geq 1, & \mbox{\small for all sets X s.t. $|X \cap T|$ is odd,} \label{eq1} \\
 0 \leq x(e) \leq 1 & \mbox{for all edges $e$.} \label{eq2}
\end{eqnarray}

This implies that any feasible solution $x$ to the system of
inequalities above can be written as a convex combination of upper
$T$-joins, i.e.  $x=\sum \alpha_i v(J_i)$ where $0 \leq \alpha_i
\leq 1$ and $\sum_i \alpha_i =1$. In particular this implies that
\begin{cor}
\label{th:tjoin} If all the edge weights $w(e)$ are non-negative,
then given any feasible assignment $x(e)$ satisfying the
inequalities above, there exists a $T$-join  with  cost at most
$\sum_e w(e) x(e)$.
\end{cor}

We are now ready to prove Theorem \ref{th:main}.

\begin{lemma}
\label{l:bythree} Let $G$ be a simple (with no multiple edges
between the same pair of vertices) planar graph with weights
$|c_{uv}|$ on the edges, and let $G^*$ be its dual graph. For any
subset of vertices $T$ of $G^*$ such that $|T|$ is even, the minimum
weighted $T$-join has weight at most $(\sum_{(u,v)} |c_{uv} |)/3 $.
\end{lemma}
{\bf Proof}: Each cut-set $J$ of  $G^*$ corresponds to a cycle in
$G$. Since $G$ is simple, each cycle has length at least $3$, and
hence each cut-set $J$ of $G^*$ contains at least 3 edges. Consider
the assignment $x(e)=1/3$. It clearly satisfies Eq. (\ref{eq2}).
Moreover it also satisfies Eq. (\ref{eq1}) as $\delta(X) \geq 3$ for
all $X \subset V$, $X \neq V$. The result then follows from
Corollary \ref{th:tjoin}.
\begin{flushright}
$\Box$
\end{flushright}

Lemma \ref{l:bythree} implies Theorem \ref{th:main} immediately,
since the value of the optimal assignment is $-W$ plus twice the
weight of the optimal $T$-join which is at most $-W + 2 W/3 = -W/3$.
Observe that Theorem \ref{th:main} is tight, as seen from the
example where $G$ is a triangle with edge weights $+1,+1$ and $-1$.
Here $W=3$, but the optimal spin assignment has value $-1$. The
condition that $G$ is simple is necessary. Otherwise, consider the
graph on two vertices with two edges, one with weight $-1$ and other
with weight $+1$. Here $W=2$, but the optimal assignment has value $0$.

\subsection{The Approximation Algorithm}
\label{sec:approx}

Given Theorem \ref{th:main}, the PTAS follows using some ideas in
\cite{kha96}. We begin by describing the notions of $p$-outerplanar
graphs and tree-widths. An outerface of a planar graph drawn in the
plane is the set of edges that constitute the boundary of the
drawing. For a tree, the outerface is the set of all the edges.
An outerplanar or 1-outerplanar graph is a planar graph that has an
embedding in the plane with all vertices appearing on the outerface.
Hence examples of 1-outerplanar graphs are trees or the graph
consisting of two cycles that share a common vertex. One can define
a $p$-outerplanar graph recursively as (see the example in Figure
\ref{fig:oplanar}):

\begin{defn}[$p$-outerplanar graphs]
A $p$-outerplanar graph is a planar graph that has an embedding in
the plane such that removing all the vertices on the outer face
gives a $(p-1)$-outerplanar graph.
\end{defn}

\begin{figure}[htb]\begin{center}\includegraphics[scale=.15]{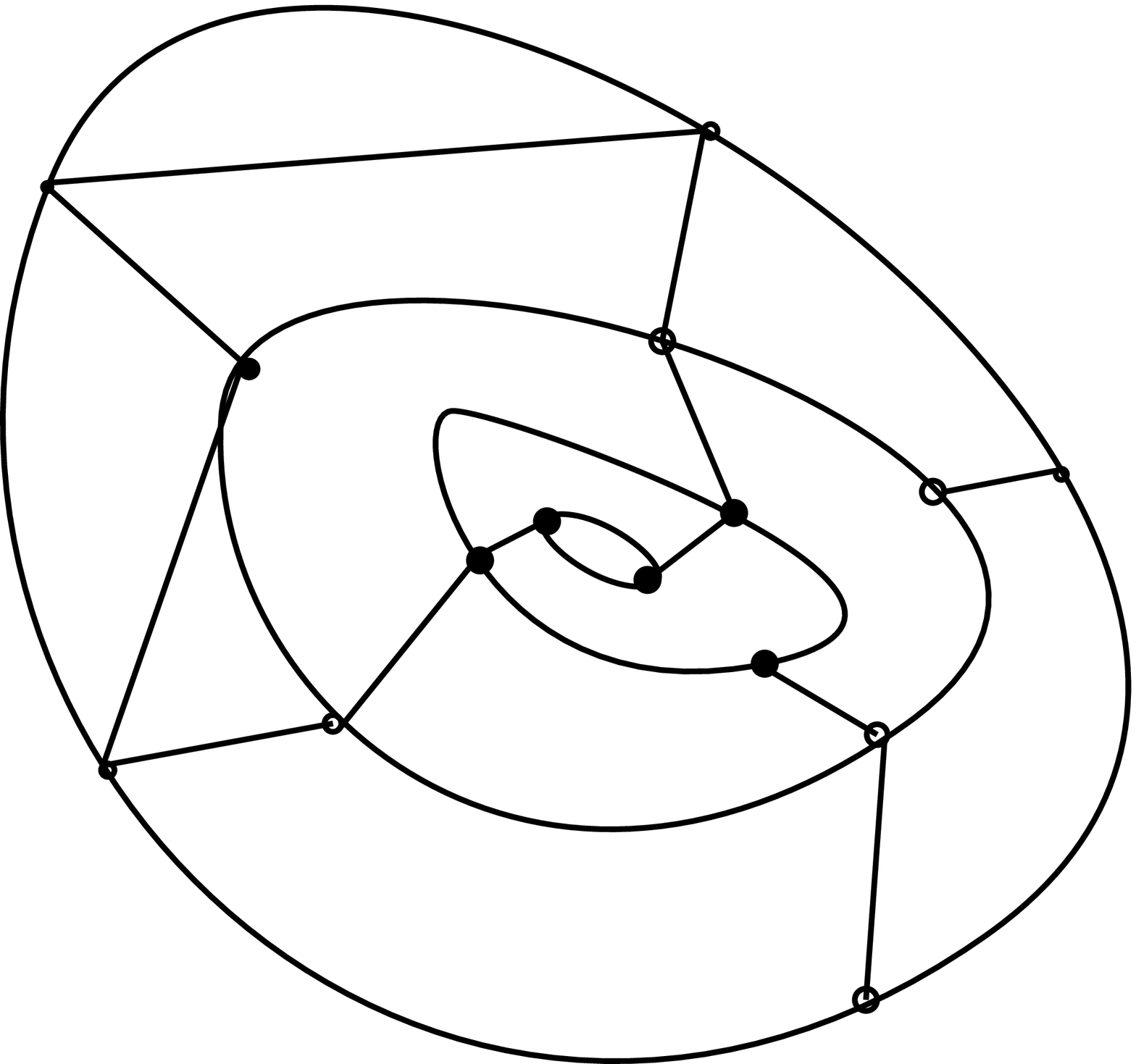}\end{center}
\fcaption{\label{fig:oplanar}A planar graph which is drawn as a
4-outerplanar graph.}\end{figure}


The notion of tree decompositions (TD) was introduced by
Robertson and Seymour \cite{RS86},
see also \cite{kha96,bod93}.  Roughly speaking, a TD allows one to map
the Ising spin Hamiltonian $H(S)$ on a graph $G=(V,E)$ to a new spin Hamiltonian
$H_{tree}(\Theta)$ that depends on  spins $\Theta_t\in \{0,1\}^b$ living at vertices $t$ of some tree $T$.
 The spin $\Theta_t$ represents the ``opinion" that a vertex $t$ has about value of
spins $S_u$ in some subset of vertices $B_t\subseteq V$ called a
{\it bag}. Accordingly, $\Theta_t$ may take $2^b$ values, where $b$
is the number of spins in $B_t$. Suppose we can
form
$m$ bags $B_1,\ldots,B_m\subseteq V$ associated with the $m$
vertices of $T$ such that (i) every vertex $u\in V$ is contained in
some bag $B_t$; (ii) for every edge $(u,v)\in E$, some bag $B_t$
contains both $u$ and $v$; (iii) a set of bags containing any given
vertex of $G$ forms a subtree of $T$. These data specify a TD of $G$
with a size $m$ and a {\em tree-width} $b-1$. The rules (i),(ii)
guarantee that one can distribute the terms $c_{uv} S_u S_v$ and
$d_u S_u$ of the Hamiltonian $H(S)$ over the bags $B_1,\ldots,B_m$
such that every term appears in exactly one bag. This distribution
defines Hamiltonians $H_{tree,t}(\Theta_t)=\sum_{(u,v)\in B_t}
c_{uv} S_u S_v + \sum_{u\in B_t} d_u S_u$, where the spins $S_u$ are
determined by $\Theta_t$. Define $H_{tree}(\Theta)=\sum_{t=1}^m
H_{tree,t}(\Theta_t)$. For every edge $(s,t)\in E(T)$ let us say
that $\Theta_s$ and $\Theta_t$ are consistent on $(s,t)$ iff their
opinions about any spin $u\in B_s\cap B_t$ agree. The rule (iii)
guarantees that $\Theta$ is consistent with some spin configuration
$S$ iff   $(\Theta_s,\Theta_t)$ are consistent on every edge of $T$.
Accordingly, $\min_S H(S) = \min_\Theta' H_{tree}(\Theta)$, where
$\min'$ means that the consistency condition on every edge of $T$ is
imposed. The optimal solution $\Theta$ can be found very efficiently
using the standard dynamic programming approach since the problem is
defined on a tree. It requires a running time of $O(m 4^{b})$,
see~\cite{bod93}.

It is known that a $p$-outerplanar graph has a TD with a size
$m=2n-1$ and a tree-width at most $3p-1$. Such a TD can be computed
in time $O(pn)$, see~\cite{Bodlaender98}. Summarizing, the minimum
energy of the Ising spin glass Hamiltonian on a $p$-outerplanar
graph with $n$ vertices can be found in time $T=O(n2^{6p})$.

Our algorithm works as follows. Given the planar graph $G$, one first
constructs a drawing of $G$ in the plane. This can be done in linear
time, using for example the algorithm of Hopcroft and Tarjan
\cite{HT74}. This gives an outerplanar decomposition of $G$, see
e.g. Fig \ref{fig:oplanar}. Say it is $h$-outerplanar ($h$ could be
as large as $O(n)$). Partition the vertices into levels
$V_1,\ldots,V_h$ where $V_1$ is the outer face and $V_i$ is the
outer face obtained by removing $V_1,\ldots,V_{i-1}$.  Let $E_i$ be
a set of edges that go from $V_{i}$ to $V_{i+1}$. For
$j=0,\ldots,t-1$,  let $G_j$ be the union  of sets $E_i$ for all $i = j \, (\mbox{modulo}\,  t)$. (Recall that $t\equiv 1/\epsilon$.)
 As each edge
lies in at most one set $G_j$,  there exists some index $j$ such
that the sum of $|c_{uv}|$ over all edges in $G_j$ is at most
$\epsilon C$. Remove all the edges in $G_j$ from the graph $G$. This
decomposes $G$ into a disjoint collection of
$t$-outerplanar graphs
$F_1,F_2,\ldots, F_{\epsilon h}$.
 We find the minimum energy
separately on each of these subgraphs.

Now consider the quality of the solution obtained for the
decomposed problem.
Let
$H_{sub,j}(S)=H(S)-\sum_{(u,v) \in G_j} c_{uv} S_u S_v$ and let the
optimal solution for $H_{sub,j}$ be $S_{opt}'$. By the
reasoning above there exists $j$ such that
\be \label{error} H_{sub,j}(S_{opt}') \leq  H_{sub,j}(S_{opt}) \leq
H(S_{opt})+\epsilon W. \ee Furthermore, $H(S_{opt}') \leq H_{
sub,j}(S_{opt}')+\epsilon W$. Thus $H(S_{opt}') \leq H(S_{opt})+2
\epsilon W \leq (1-6\epsilon) H(S_{opt})$ by Theorem \ref{th:main}.
It follows that we can get a PTAS with a relative error $6\epsilon$ by
trying all possible $j=0,\ldots,t-1$ and choosing the optimal solution $S_{opt}'$
that yields the smallest value of $H(S_{opt}')$.

For a fixed $j$ finding $S_{opt}'$ requires time $T_j=
\sum_{a=1}^{\epsilon h} O(|F_a| 2^{6t})=O(\epsilon n 2^{6t})$.
Thus the overall running time of the PTAS is $T=\sum_{j=0}^{t-1} T_j =O(n2^{{6}/{\epsilon}})$.
Choosing $\delta = 6 \epsilon$ this implies
that the algorithm obtains an assignment with the energy at most $
(1-\delta)H(S_{\rm opt})$ in time $O( n 2^{36/\delta})$.


\section{Quantum Ising Spin Glass}
\label{sec:quantum}

The following theorem is the quantum equivalent of Theorem
\ref{th:main} and will be instrumental in proving our results:

\begin{thm}
The minimum eigenvalue $\lambda(H)$ of $H$ for a planar graph
$\lambda(H)\le -\sum_u ||L_u||/5-W/(5\cdot 3^5)$,
where $W=\sum_{(u,v)\in E} \| Q_{u,v}\|$. \label{thm:extq}
\end{thm}

{\bf Proof}: The strategy will be to upper bound $\lambda(H)$ by
$\lambda_{sep}(H)$, where $\lambda_{sep}$ is the minimal energy
achieved on tensor products of states
$|0\ra,|1\ra,|+\ra,|-\ra,|+i\ra,|-i\ra$, where
\[
|\pm \ra =(|0\ra\pm |1\ra)/\sqrt{2}, \quad |\pm i\ra =(|0\ra \pm i
\, |1\ra)/\sqrt{2}.
\]
It is enough to prove that $\lambda_{sep}$ is an extensive quantity
and this can be achieved using the classical result, Theorem
\ref{th:main}. Let us first prove the Theorem for the special case
then $L=0$, that is $H=Q$ involves only interactions quadratic in
Pauli operators. For every edge $(u,v)\in E$ the interaction
$Q_{u,v}$ generally involves all $9$ combinations of Pauli
operators. We will choose one of them that has the largest magnitude
and call it a {\it dominating coupling} (ties are broken arbitrarily).
 For example, if
$Q_{u,v}=3X_u\otimes Y_v -4 Z_u\otimes X_v$, then the edge $(u,v)$
has dominating coupling $-4 Z_u\otimes X_v$. We have

\begin{lemma}
\label{lemma:dom} Suppose $Q_{u,v}$ has a dominating coupling
$c_{uv} \, P^\alpha_u\otimes P^\beta_v$, where $P^\alpha,P^\beta\in
\{X,Y,Z\}$. Then \be \label{c>=} |c_{uv}|\ge \frac19 \|Q_{u,v}\|.
\ee
\end{lemma}
{\bf Proof:} Indeed, otherwise the triangle inequality would imply
$\|Q_{u,v}\| \le 9 |c_{uv}| <\|Q_{u,v}\|$.
\begin{flushright}
$\Box$
\end{flushright}
We shall now partition the edges $E$ into several subsets $E=\cup_j
E_j$, such that the dominating couplings in each subset $E_j$
commute with each other, that is, the sum of dominating couplings
over $E_j$ is equivalent to a classical Ising Hamiltonian up to a
local change of basis. First of all, since $G$ is a planar graph we
can color its vertices by $\{1,2,3,4\}$ such that adjacent vertices
have different colors. A map $f\, :\, V\to \{X,Y,Z\}$ that assigns a
Pauli operator to every vertex of $G$ will be called a {\it Pauli
frame} if $f(u)$ depends only on a color of $u$. Consider the
following Pauli frames.
\begin{center}
\begin{tabular}{|cccc|}
\hline
1 & 2 & 3 & 4 \\
\hline
X & X & X & X\\
X & Y & Z & Y \\
X & Z & Y & Z\\
Y & X & Z & Z\\
Y & Y & Y & X\\
Y & Z & X & Y \\
Z & X & Y & Y \\
Z & Y & X & Z \\
Z & Z & Z & X\\
\hline
\end{tabular}
\end{center}
This table forms an orthogonal array of strength two with alphabet
$\{X,Y,Z\}$, that is every pair of columns contains every possible
combination of two Pauli operators exactly one time. Let
$f_1,\ldots,f_9$ be the corresponding Pauli frames. Denote $E_j$ a
subset of edges $(u,v)\in E$ such that $(u,v)$ has a dominating
coupling
\[
c_{uv}\, P^{f_j(u)}_u\otimes P^{f_j(v)}_v.
\]
Then we conclude that \be \label{E_j} E_j \cap E_k =\emptyset \quad
\mbox{if} \quad j\ne k, \quad \mbox{and} \quad \cup_{j=1}^9 E_j =E.
\ee For every Pauli frame $f_j$ we can introduce a classical Ising
Hamiltonian $Q_j$ obtained from $Q$ by restricting the whole Hilbert
space to classical states in the Pauli frame $f_j$ (that is, if
$f_j(u)=X$ for some qubit $u$, we allow this qubit to be in either
of states $|+\ra$ and $|-\ra$; if  $f_j(u)=Z$, we allow  $u$ to be
in either of states $|0\ra$, $|1\ra$, e.t.c.). By definition, \be
\label{Q<=Q_j} \lambda(Q)\le \lambda(Q_j). \ee Note that for every
edge $(u,v)\in E_j$ the dominating coupling in $Q_{u,v}$ is diagonal
in the Pauli frame $f_j$. Thus, applying Theorem~\ref{th:main} to
$Q_j$ we obtain \be \label{C_j} \lambda(Q_j) \le -\frac13
\sum_{(u,v)\in E_j} |c_{uv}| \le -\frac1{3^3} \sum_{(u,v)\in E_j} \|
Q_{u,v}\|, \ee where the second inequality follows from
Lemma~\ref{lemma:dom}. Combining
Eqs.~(\ref{E_j},\ref{Q<=Q_j},\ref{C_j}) we arrive to \be
\lambda(Q)\le \frac19 \sum_{j=1}^9 \lambda(Q_j) \le - \frac1{3^5}
\sum_{(u,v)\in E} \|Q_{u,v}\|. \ee It remains to generalize this
bound to the case $L\ne 0$. We can show that, similar as in the
classical case (see Claim \ref{cl1}), the following holds:\\

\begin{lemma}
\label{lemma:L} One can choose a ground-state $|\psi_0\ra$ of $Q$
such that $\la \psi_0|L|\psi_0\ra\le 0$.
\end{lemma}
{\bf Proof:} Consider an anti-unitary operator $W$ (known as
Kramers-Wannier duality) such that
\[
W\,|\phi\ra = \bigotimes_{u\in V} Y_u\, |\phi^*\ra,
\]
where the complex conjugation is performed in $|0\ra,|1\ra$ basis.
One can check that $P^\alpha_u \, W = -W P^\alpha_u$ for any
single-qubit Pauli operator $P^\alpha_u$. It follows that
\[
Q W =W Q, \quad L W = -W L.
\]
Thus, if one defines a state $|\phi_0\ra=W\, |\psi_0\ra$, one gets
\[
\la \phi_0|L|\phi_0\ra =- \la \psi_0|L|\psi_0\ra, \quad \la
\phi_0|Q|\phi_0\ra = \la \psi_0|Q|\psi_0\ra.
\]
It follows that $|\phi_0\ra$ is also a ground state of $Q$ and  one
of the expectations values $\la \phi_0|L|\phi_0\ra$ or $\la
\psi_0|L|\psi_0\ra$ is non-positive.
\begin{flushright}
$\Box$
\end{flushright}

Now define $5$ states $\rho_1,\ldots,\rho_5$ such that\\
(i) For $j=1,\ldots,4$ a state $\rho_j$ sets qubits $u\in V$ of
color $j$ to the ground state of  $L_u$;
all other qubits are set to the maximally mixed state.\\
(ii) $\rho_5$ is a ground state of $Q$ such that $\trace(\rho_5\,
L)\le 0$.

\noindent Then one has the following inequalities \be \label{Ineq1}
\trace(Q\, \rho_j)=0 \quad \mbox{for} \quad j=1,\ldots,4, \quad
\mbox{and} \quad \trace(Q\, \rho_5) =\lambda(Q). \ee \be
\label{Ineq2} \trace(L\, \rho_j) = - \sum_{u\, :\,
{\mathrm{color}}(u)=j} \|L_u\| \quad \mbox{for} \quad j=1,\ldots,4,
\quad \mbox{and} \quad \trace(L\,\rho_5)\le 0. \ee Let $\rho$ be the
uniform probabilistic mixture of $\rho_1,\ldots,\rho_5$. Then
Eqs.~(\ref{Ineq1},\ref{Ineq2}) imply that
\[
\trace(\rho\, H) \le -\frac15 \sum_{u\in V} \|L_u\| -\frac1{5\cdot
3^5} \sum_{(u,v)\in E} \|Q_{u,v}\|. \label{eq:fullbound}
\]
\begin{flushright}
$\Box$
\end{flushright}

\subsection{The Approximation Algorithm for the Quantum Problem on a Planar Graph with Bounded Degree}
\label{sec:qapprox}

We use the following result of Klein, Plotkin and Rao \cite{kpr} on
decomposing planar graphs.
\begin{thm} Let $G =(V,E)$ be an undirected planar graph with non-negative edge weights, and let $W$
denote the total edge weight. Then, given any $\epsilon > 0$, there
is a subset of edges $E'$ with total weight at most $\epsilon W$
such the removing these edges decomposes the graph $G$ into
components each of which has weak diameter at most $c/\epsilon$ with
respect to $G$. Here $c$ is fixed constant independent of
$\epsilon$. \label{thm:kpr}
\end{thm}

Here is the definition of a weak diameter:
\begin{definition}{\bf Weak Diameter:}
Let $G=(V,E)$ be a graph and let $G'=(V',E')$ be a subgraph of $G$.
We say that $G'$ has weak diameter $d$ with respect to $G$, if for any
two vertices $v,w \in V'$, their distance in $G$ is at most $d$.
\end{definition}

If $G$ is a planar graph with maximum degree $d$, Theorem
\ref{thm:kpr} implies that each component can have at most
$d^{O(1/\epsilon)}$ vertices and hence the minimum eigenvalue
problem for a Hamiltononian $H$ restricted to every component
 can be solved
in time $2^{d^{O(1/\epsilon)}}$.


There is a linear time algorithm to determine the set of edges $E'$.
The algorithm works as follows: Let $\delta = \epsilon/3$. Root the
graph $G$ at arbitrary vertex and consider the breadth first tree. A
vertex is said to be at level $i$, if it is at distance $i$ from the
root. For $j=0,\ldots,1/\delta-1$, let $E_j$ denote the set of edges
that connect two vertices at level $i$ and $i+1$ where $i \equiv j $
(modulo $1/\delta$). Choose the set $E_j$ with least weight and
remove these edges from $G$. Let $G_1$ denote the obtained graph.
Now consider each of the components of $G_1$ and apply this
procedure again to obtain the graph $G_2$. Finally, apply the same
procedure to $G_2$ to obtain $G_3$
(one applies the procedure three times because planar graphs $K_{3,3}$ minor free).
The result of Klein, Plotkin and
Rao \cite{kpr} shows that $G_3$ has weak diameter at most
$O(1/\delta) = O(1/\epsilon)$.
Moreover the weight of edges removed
is at most $3 \delta = \epsilon$ fraction of the total weight.

Let $H=\sum_{(u,v)\in E} Q_{u,v} + \sum_{u\in V} L_u$ be a quantum Ising spin glass Hamiltonian
defined on a graph $G=(V,E)$.
Define a weight associated with an edge $(u,v)$ as $\|Q_{u,v}\|$.
Let $H_{sub}$ be a Hamiltonian obtained from $H$ by keeping all the
linear terms $L_u$ and the quadratic
terms $Q_{u,v}$ associated with edges of a subgraph $G_3$ defined above.
By definition of $G_3$ one has $\| H-H_{sub}\| \le \epsilon W$
and thus $|\lambda(H_{sub})-\lambda(H)|\le \epsilon W$.
Theorem~\ref{thm:extq} implies that $|\lambda(H_{sub}) - \lambda(H)|\le c\epsilon |\lambda(H)|$
for some numeric constant $c$. Thus one can approximate $\lambda(H)$ with any fixed
relative error $\epsilon$ in time $n 2^{d^{O(1/\epsilon)}}$.



\section{Quantum Ising Spin Glass Problem on a Star Graph}
\label{sec:qstar}
The construction of PTAS for classical Hamiltonians on
planar graphs presented in Section~\ref{sec:planar}
relies on the fact that the classical problem on a tree (or any graph of constant treewidth)
can be solved efficiently using the dynamic programming.
Unfortunately, this method does not work for quantum Hamiltonians.
In this section we develop a new technique that allows one to
obtain a PTAS for the quantum Ising spin glass Hamiltonian
on a {\it star graph}  --- a tree that consists of $n+1$ vertices with
one vertex having degree $n$ and $n$ vertices having degree $1$.
The corresponding graph is $G=(V,E)$, where $V=\{0,1,\ldots,n\}$
and $E=\{(0,1),(0,2),\ldots,(0,n)\}$.
We shall refer to spins sitting at vertices $1,2,\ldots,n$ as {\it bath spins}
and the spin sitting at the vertex $0$ as {\it central spin} (by analogy with the
central spin problem studied in condensed matter physics~\cite{Gaudin}).
Let the Hamiltonian be
 \be \label{H} H=H_0+\sum_{j=1}^n H_{0,j} \ee
 where $H_{0}$ is a linear term acting on the central spin and
$H_{0,j}$ is the interaction between the central spin and $j$-th
bath spin (which can include both quadratic and linear terms).
\begin{theorem}
\label{thm:central}
Suppose there exist constants $0<a\le b$ such that $a\le \|H_{0,j}\| \le b$ for
all $j$. Then one can approximate the smallest eigenvalue $\lambda(H)$
with a relative error $\epsilon$ in time $n^{\epsilon^{-O(1)}}$.
\end{theorem}
In the rest of this section we prove Theorem~\ref{thm:central}.
We start from proving that  $\lambda(H)$ can be computed {\it exactly} in time ${\rm
poly}(n)$ as long as the number of distinct interactions $H_{0,j}$
is bounded by a constant. We shall use
\begin{lemma}
\label{lemma:symmetry} Suppose the interactions $H_{0,j}$ are the
same for some subset of bath spins $S\subseteq \{1,\ldots,n\}$. Then
one can choose a pure ground state of $H$ symmetric under
permutations of spins in $S$.
\end{lemma}

{\bf Proof}: Without loss of generality $S=\{1,2,\ldots,k\}$. Let
$|\Psi\ra$ be a ground state of $H$. Denote $W_{j,k}$ the swap of
qubits $j$ and $k$. Assume that $|\Psi\ra$ is not symmetric under
permutations of spins in $S$. Without loss of generality,
$W_{1,2}\,|\Psi\ra \ne |\Psi\ra$. There are two cases: (i)
$W_{1,2}\, |\Psi\ra$ is proportional to $|\Psi\ra$. Then $W_{1,2}\,
|\Psi\ra = - |\Psi\ra$, since $W_{1,2}$ has eigenvalues $\pm 1$.
Therefore $|\Psi\ra=|\Psi^-\ra_{12} \otimes |\Psi_{else}\ra$, where
$|\Psi^-\ra=\frac{1}{\sqrt{2}}(\ket{01}-\ket{10})$ is the singlet
state and $|\Psi_{else}\ra$ is some state of spins
$\{0,3,4,\ldots,n\}$. The second case is (ii) $W_{1,2}\, |\Psi\ra$
and $|\Psi\ra$ are linearly independent. Then the anti-symmetrized
state $|\Psi'\ra=|\Psi\ra - W_{12}\, |\Psi\ra$ is non-zero. On the
other hand, $|\Psi'\ra$ is a ground state of $H$ since $W_{12}$
commutes with $H$. We conclude that
$|\Psi'\ra=|\Psi^-\ra_{12}\otimes |\Psi_{else}\ra$. In both case we
conclude that $H$ has a ground state
$|\Psi\ra=|\Psi^-\ra_{12}\otimes |\Psi_{else}\ra$.

Since the energy of a state depends only upon the reduced density
matrices $\rho_{0,j}$, we can replace the antisymmetric singlet
$|\Psi^-\ra$ by the symmetric EPR state $|\Psi^+\ra$ without
changing the energy. On the other hand, any state with energy equal
to the ground-state energy must be a ground state. We conclude that
$H$ has a ground-state $|\Psi\ra=|\Psi^+\ra_{12}\otimes
|\Psi_{else}\ra$.

By iterating the arguments above one concludes that $H$ has a ground
state \be \label{Psi=} |\Psi\ra = |\Psi^+_{12}\ra\otimes \cdots
\otimes |\Psi^+_{k-1,k}\ra \otimes |\Psi_{else}\ra\equiv |\Psi_S\ra
\otimes |\Psi_{else}\ra. \ee where $|\Psi_{else}\ra$ is some state
of all spins $j\notin S$ (if $k$ is odd then there will be one
unpaired spins in $S$; this will not change the arguments below very
much). Now we can symmetrize $|\Psi\ra$ by brute force method. Let
\[
\Pi = \frac1{k!}\sum_{\tau\in S_k} W(\tau)
\]
be the projector onto the symmetric subspace, where $W(\tau)$ is the
unitary operator implementing a permutation $\tau$ of $k$ spins in
$S$. Note that the state $|\Psi_S\ra$ in Eq.~(\ref{Psi=}) has
non-negative amplitudes in the standard basis. Therefore $W(\tau)\,
|\Psi_S\ra$ also has non-negative amplitudes. Therefore $\Pi\,
|\Psi_S\ra \ne 0$, and, accordingly, $|\Psi'\ra=\Pi_S\otimes
I_{else}\, |\Psi\ra$ is a non-zero state symmetric under
permutations of spins in $S$. On the other hand, since $W(\tau)$
commutes with $H$, $|\Psi'\ra$ is a ground state of $H$.
\begin{flushright}
$\Box$ \end{flushright}

This result implies that we can look for a ground state that
``occupies" only a $|S|+1$ dimensional subspace of the $2^{|S|}$
dimensional Hilbert space describing spins in $S$. If we have $M$
distinct interactions, the dimension of the space in which the
optimization takes place is $\Pi_{i+1}^M (|S_i|+1) \leq n^M$ which
is polynomial in $n$. Thus the optimization problem for constant $M$
can be solved efficiently in $n$.

In order to map the general problem onto one in which we have
constant number of distinct interaction, we apply a coarse-graining
procedure to the general Hamiltonian Eq.~(\ref{H}). One can show

\begin{lemma}
\label{lemma:cgrain} For any $0<a<1$ define a set $M_{a}$ of
$2$-qubit Hamiltonians $H$ satisfying $a\le \|H\| \le 1$. For any
$\epsilon>0$ there exist $2$-qubit Hamiltonians $G_1,\ldots,G_M$,
$M=O((a \epsilon)^{-32})$ such that $ \min_{\alpha} \| G_\alpha -
H\| \le \epsilon\, \|H\|$ for any $H\in M_{a}$.
\end{lemma}

{\bf Proof}: It is enough to satisfy $\min_{\alpha} \| G_\alpha -
H\| \le \epsilon a$. A $2$-qubit Hamiltonian satisfying $\|H\|\le 1$
lives in a $2\times 2\times \cdots \times 2$ cube in $\RR^{32}$.
Construct $\epsilon a$-mesh, count the number of points.
\begin{flushright}
$\Box$
\end{flushright}

Now we are ready to prove Theorem~\ref{thm:central}.
Without loss of generality, $b=1$ (otherwise multiply $H$ by $b^{-1}$).
Applying Lemma~\ref{lemma:cgrain} to every interaction $H_{0,j}$
one can partition the $n$ bath spins into  $M=O((a \epsilon)^{-32})$
subsets $S_1,\ldots,S_M$ such that $\|H_{0,j} - G_\alpha\|\le
\epsilon\, \|H_{0,j}\|$ for all $j\in S_\alpha$. We define a
coarse-grained Hamiltonian $\tilde{H}=H_{0} + \sum_{\alpha=1}^M
\sum_{j\in S_\alpha} G_{\alpha}[j].$ Here the notation $G_{\alpha}[j]$ means that
$G_{\alpha}$  acts on the spins $0$ and $j$. We have $\|H -\tilde{H}\| \le
\epsilon \sum_{j=1}^n \|H_{0,j}\| \leq \epsilon 5 \cdot 3^5
|\lambda(H)|$ where the second inequality follows from  Theorem \ref{thm:extq}.
Therefore, $|\lambda(H)-\lambda(\tilde{H})|\le c\epsilon |\lambda(H)|$ for
some numeric constant $c$.
  The classical PTAS will
find the ground-state (that is, a poly(n)-sized classical
description of this state) and the ground-state energy
$\lambda(\tilde{H})$ of the coarse-grained Hamiltonian.
 Lemma~\ref{lemma:symmetry} implies that it requires
 time $n^{\epsilon^{-O(1)}}$.

\section{Discussion}
\label{sec:dis}

An important open question is whether there exists a classical or
quantum PTAS for the general quantum Ising spin glass problem on
planar graphs. It is clear that some new techniques will be needed
to settle this problem. A simpler problem in this realm would the
quantum Ising spin glass problem on a tree with unbounded degree.
Note that even in the simplest case of star graphs the existence of PTAS
for the quantum problem is not proved (Theorem~\ref{thm:central} assumes
the constant lower and upper bounds on the norm of interactions $H_{0,j}$).

One interesting approach to address these quantum problem may be to
consider quantum or classical algorithms that output the thermal state
$e^{-H/T}/Z$ at temperature $T$ for the Ising spin glass problem. Such thermal state will typically provide
a PTAS for the ground-state energy problem. One
can prove this by showing that the average energy $\langle
H\rangle_T={\rm Tr}H e^{-H/T}{Z}$ is bounded as $|\langle
H\rangle_T-\lambda(H)| \leq 2 nT$. This bound follows from the fact
that for the free-energy $F(T)$ we have $|F(T)-\langle H\rangle_T|
\leq nT$ and $|F(T)-\lambda(H)|=|F(T)-F(0)|\leq nT$. When the
ground-state energy $\lambda(H)$ scales with $n$ (e.g. for bounded
weights $|c_{uv}| \leq c, ||Q_{uv}|| \leq c$ one gets this from Theorems
\ref{th:main} and \ref{thm:extq}), the error in the approximation can be made $\epsilon\lambda(H)$ for $T=O(\epsilon)$. This also shows
that finite but small temperature implementation of adiabatic
quantum computation will generally provide a PTAS-approximation to
the ground-state energy problem (assuming that, say, for
bounded-degree graphs beyond the planar ones, the ground-state
energy will be extensive, scales with $n$).

The difference between classical and quantum behavior on tree
graphs is also witnessed by the fact that the algorithm of classical
belief propagation (for zero temperature this essentially
corresponds to dynamic programming) converges efficiently on trees,
whereas quantum belief propagation will only work when additional
conditions are fulfilled \cite{QBP:poulin_leifer,QBP:hastings}. It
is expected that for bounded-degree trees the quantum belief
propagation algorithm of \cite{QBP:hastings} at finite temperature
$T$ will give rise to a PTAS.

\section*{Acknowledgements}
SB and BMT acknowledge support by NSA and ARDA through ARO contract
number W911NF-04-C-0098.

\bibliographystyle{hunsrt}

\end{document}